# *RISCBench:* Benchmarking RISC-V Orchestration Efficiency in FPGA and FPGA-Like Computing Engines

## An Industry Evaluation of Control-Plane Bottlenecks and Sustained Throughput Metrics

Dave Ojika[†]
Flapmax
Tampa, FL, USA
dave@flapmax.com

Projjal Gupta
University of Florida
Gainesville, FL, USA
projjalgupta@ufl.edu

Preethi Budi[*]
Flapmax
Gainesville, FL, USA
preethibudi@flapmax.com

Herman Lam
NSF Center for Space, High-Performance & Resilient Computing
University of Florda
Gainesville, FL, USA
hlam@ufl.edu

Shreya Mehrotra
Altera Corporation
Austin, TX, USA
shreya.mehrotra@altera.com

## Introduction

Heterogeneous computing systems increasingly rely on lightweight control-plane cores to orchestrate data movement, synchronization, and scheduling across accelerators and reconfigurable fabrics. RISC-V has emerged as a widely adopted orchestration substrate, spanning soft FPGA cores and hard accelerator-class cores. Conventional metrics such as FLOPs, TOPS/W, and energy per operation primarily characterize arithmetic capability, but provide limited visibility into orchestration efficiency [1], even though orchestration behavior often determines sustained throughput. This limitation becomes more pronounced as heterogeneous systems move toward tighter integration while control-plane bottlenecks persist.

We introduce RISCBench, a lightweight benchmark suite and open methodology designed to quantify orchestration efficiency. A key component of this methodology is the Sustained Instantaneous Throughput (SIT) metric, which accumulates the contribution of high-efficiency execution to realized sustained throughput. In contrast to peak-oriented descriptors, SIT reflects how coordination, scheduling, and data placement affect the persistence of high-efficiency execution under realistic orchestration conditions. By emphasizing sustained behavior rather than instantaneous peak rates, RISCBench and SIT expose orchestration inefficiencies not captured by conventional peak-oriented descriptors, enabling more meaningful evaluation of heterogeneous FPGA and accelerator-class computing systems.

[†]Corresponding author. Email dave@flapmax.com
[*]Work done while Preethi Budi was a research intern at Flapmax and affiliated with the University of Florida.

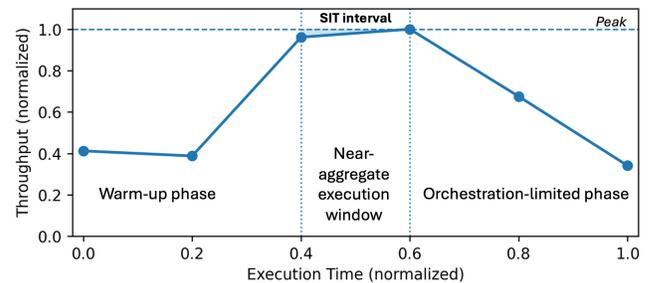

Fig. 1. Normalized throughput over execution time showing a brief near-aggregate execution window followed by orchestration-induced degradation; SIT captures the accumulated high-efficiency execution.

## Experimental Snapshot

RISCBench was prototyped using a Nios-V soft RISC-V core on an FPGA to validate integration within reconfigurable environments [2], and evaluated on an accelerator-class platform embedding large numbers of hard RISC-V control cores dedicated to orchestration [3]. SRAM-resident tiled matrix kernels were used to minimize external memory effects and expose orchestration behavior. Execution traces (Fig. 1) show near-aggregate throughput in early execution windows, followed by degradation as synchronization overhead and memory residency transitions accumulate. Peak throughput is observed under full on-chip residency, while sustained performance declines with increasing coordination and bandwidth contention. These observations indicate that, as heterogeneous integration advances toward more tightly coupled accelerator architectures, orchestration behavior increasingly shapes realized throughput. SIT provides a platform-independent descriptor of sustained execution behavior that complements peak-based metrics, including AI inference applications. To support reproducibility and broader adoption, RISCBench is released as an open-source benchmark suite at www.riscbench.com, inviting contributions from the chip architecture, EDA research, and systems communities.